\documentclass[12pt]{article}

\usepackage{amsmath}
\usepackage{amssymb}
\usepackage{amstext}
\usepackage{amsopn}
\usepackage{amsfonts}
\usepackage{amsxtra}
\usepackage[english]{babel}
\usepackage{graphicx}
\usepackage{float}
\usepackage{bm}
\usepackage{multirow}
\usepackage{hyperref}
\usepackage{lineno}
\usepackage{color}
\usepackage[superscript]{cite}
\usepackage{times}
\usepackage{fullpage}

\makeatletter
\renewcommand\@biblabel[1]{#1.}
\makeatother

\newcommand{\icm}{\ensuremath{\textrm{cm}^{-1}}}

\newcommand{\LHN}{LuH$_{2 \pm x}$N$_{y}$}
\newcommand{\EF}{$E_{\text{F}}$}

\begin{document}


\begin{trivlist}
  \item[] {\Large\textsf{\textbf{Pressure-induced color change arising from transformation between intra- and inter-band transitions in LuH$_{2\pm x}$N$_{y}$}}}
\end{trivlist}


\begin{trivlist}
  \item[] Zhe~Liu$^{1,\ast}$, Yingjie~Zhang$^{1,\ast}$, Shenyang~Huang$^{2,\ast}$, Xue Ming$^{1}$, Qing~Li$^{1}$, Chenghao~Pan$^{3}$, Yaomin~Dai$^{1,\dag}$, Xiaoxiang~Zhou$^{1}$, Xiyu~Zhu$^{1}$, Hugen~Yan$^{3,\dag}$, \& Hai-Hu~Wen$^{1,\dag}$


 \item[] $^{1}$\emph{National Laboratory of Solid State Microstructures and Department of Physics, Collaborative Innovation Center of Advanced Microstructures, Nanjing University, Nanjing 210093, China}
 \item[] $^{2}$\emph{Institute of Optoelectronics, Fudan University, 200433 Shanghai, China}
 \item[] $^{3}$\emph{State Key Laboratory of Surface Physics, Key Laboratory of Micro and Nano-Photonic Structures of the Ministry of Education, and Department of Physics, Fudan University, 200433 Shanghai, China}


  \vspace{2mm}
  \item[] $^{\ast}$These authors contributed equally to this work.
  \item[] $^{\dag}$email: ymdai@nju.edu.cn; hgyan@fudan.edu.cn; hhwen@nju.edu.cn
  \vspace{4mm}
\end{trivlist}


\boldmath
\begin{trivlist}
\item[] {\bf The pressure-induced color change in the nitrogen-doped lutetium hydride has triggered extensive discussions about the underlying physics. Here, we study the optical response of LuH$_{2 \pm x}$N$_{y}$ in a broad frequency range at ambient pressure and its evolution with pressure in the visible spectral range. The broad-band optical spectra at ambient pressure reveal a Drude component associated with intra-band electronic transitions and two Lorentz components (L1 and L2) arising from inter-band electronic transitions. The application of pressure causes a spectral weight transfer from L1 to the Drude component, leading to a blue shift of the plasma edge in the reflectivity spectrum alongside a reduction of the high-frequency reflectivity. Our results suggest that the pressure-induced color change in LuH$_{2 \pm x}$N$_{y}$ is closely related to the transformation between intra- and inter-band electronic transitions, providing new insights into the mechanism of the pressure-induced color change in LuH$_{2 \pm x}$N$_{y}$.}
\end{trivlist}
\unboldmath


\section*{Introduction}
Hydrogen-rich compounds have received enormous attention~\cite{Drozdov2015Nature,Einaga2016NP,Drozdov2019Nature,Liu2017PNAS,Peng2017PRL,Sun2019PRL,Cataldo2021PRB,Zhang2022PRL,Somayazulu2019PRL,Chen2021PRL,Hong2020CPL,Kong2021NC}, as theory has predicted that hydrogen atoms in these materials provide high-frequency phonons and strong electron-phonon coupling, which, according to the Bardeen-Cooper-Schrieffer theory of conventional superconductivity, support the emergence of high-temperature superconductivity~\cite{Ashcroft2004PRL}. Following this strategy, a series of hydrogen-rich materials have been predicted~\cite{Liu2017PNAS,Peng2017PRL,Sun2019PRL,Cataldo2021PRB,Zhang2022PRL}, and high-temperature superconductivity has been achieved in some of these materials including H$_{3}$S with $T_{c}$ = 203~K at 155~GPa~\cite{Drozdov2015Nature,Einaga2016NP}, LaH$_{10}$ with $T_{c}$ = 250--260~K at 170--180~GPa~\cite{Drozdov2019Nature,Somayazulu2019PRL,Hong2020CPL}, and YH$_{9}$ with $T_{c}$ = 243~K at 201~GPa~\cite{Kong2021NC}.

The recent report of room-temperature superconductivity at near-ambient pressure in a nitrogen-doped lutetium hydride (Lu-H-N)~\cite{Dasenbrock-Gammon2023Nature} has aroused a flurry of excitement. However, suspicions immediately arise because several groups have soon repeated the experiment independently, but found no evidence of superconductivity at near-ambient pressure~\cite{Shan2023CPL,Ming2023arXiv,Xing2023arXiv,Cai2023arXiv}. Shan et al. have reported the absence of superconductivity in LuH$_{2}$ down to 1.5~K under pressures up to 7.7~GPa~\cite{Shan2023CPL}; Ming et al. have successfully synthesized \LHN\ but revealed no superconductivity down to 10~K under pressures up to 6.3~GPa~\cite{Ming2023arXiv}; Xing et al. have carried out systematic resistance measurements on \LHN\ at temperatures from 1.8 to 300~K and pressures from 0.4 to 30~GPa, detecting no signal of superconductivity~\cite{Xing2023arXiv}. Cai et al. have studied Lu-H-N prepared from the lutetium foil and H$_{2}$/N$_{2}$ gas mixture, the same method as that reported in Ref.~\cite{Dasenbrock-Gammon2023Nature}, yet found no evidence of superconductivity~\cite{Cai2023arXiv}. Li et al. have observed a superconducting transition with $T_{c}$ = 71~K at 218~GPa in the Lu$_{4}$H$_{23}$ phase with a $Pm\overline{3}n$ symmetry~\cite{Li2023SCPMA}, which is completely different from the reported Lu-H-N system.

Despite the negative results in confirming the room-temperature superconductivity, the pressure-induced color change from blue to red in the Lu-H-N system has been consistently reproduced~\cite{Shan2023CPL,Zhang2023arXiv,Xing2023arXiv,Zhao2023arXiv}, attracting considerable interest in both experimental and theoretical communities~\cite{Zhao2023arXiv,Kim2023arXiv,Liu2023arXiv,Tao2023arXiv}. Pressure-induced color change has also been observed in YH$_{3}$~\cite{Wijngaarden2000JAC}. While Zhao et al. have revealed that the pressure-induced color change in LuH$_{2}$ is caused by a shift of the plasma edge in the reflectivity spectrum~\cite{Zhao2023arXiv}, the microscopic mechanism of the color change in the Lu-H-N system is currently under hot discussion~\cite{Kim2023arXiv,Liu2023arXiv,Tao2023arXiv}.

Here, we present a combined study on the broad-band optical response of LuH$_{2 \pm x}$N$_{y}$ at ambient pressure and the evolution of optical properties with pressure in the visible spectral range. The broad-band reflectivity and optical conductivity at ambient pressure reveal a single Drude component arising from intra-band electronic transitions and two Lorentz components (L1 and L2) originating from inter-band electronic transitions. With increasing pressure, the Drude weight increases, leading to a blue shift of the plasma edge in the reflectivity spectrum; the weight of L1 diminishes, resulting in a reduction of the high-frequency reflectivity. These observations suggest that the pressure-induced color change in LuH$_{2 \pm x}$N$_{y}$ is intimately linked to the transformation between intra- and inter-band electronic transitions, thus shedding new light on the mechanism of the pressure-induced color change in the Lu-H-N system.

\section*{Results}
%
\subsubsection*{Pressure-induced color change and visible-range reflectivity.}

%
%

Figure~1a displays the optical microscope images of our \LHN\ sample in a symmetric diamond anvil cell (DAC) with 400~$\mu$m culets at different pressures up to 33.2~GPa. As the pressure rises, the color of the sample changes from blue to red, in agreement with previous results~\cite{Dasenbrock-Gammon2023Nature,Zhang2023arXiv,Shan2023CPL,Xing2023arXiv,Zhao2023arXiv}. Note that the pressure threshold for the color change varies for samples grown by different groups, which may be ascribed to a difference in the content of N or H.

Figure~1b shows the reflectivity $R(\omega)$ of \LHN\ in the visible range at different pressures from 0 to 33.26~GPa. At ambient pressure, as shown by the blue curve, a plasma edge, i.e. a dramatic drop in $R(\omega)$, is observed near 12\,000~\icm. Such a plasma edge in $R(\omega)$ represents the fingerprint of metals~\cite{Dressel2002,Tanner2019}. $R(\omega)$ at 20\,833~\icm\ ($\lambda$ = 480~nm, blue light) is higher than the value at 14\,815~\icm\ ($\lambda$ = 675~nm, red light), suggesting that the crystal reflects more blue light than the red. This naturally accounts for the blue color of the \LHN\ compound at ambient pressure. With increasing pressure, the plasma edge shifts to higher frequency and $R(\omega)$ above 18\,000~\icm\ decreases, which is similar to previous optical measurements on LuH$_{2}$~\cite{Zhao2023arXiv}. In Fig.~1c, we plot the values of $R(\omega)$ for red light $R_{\text{red}}$ (red solid circles) and blue light $R_{\text{blue}}$ (blue solid circles) as a function of pressure. It is noticeable that $R_{\text{red}}$ rises and $R_{\text{blue}}$ drops with increasing pressure, well accounting for the pressure-induced color change from blue to red in \LHN.

\subsubsection*{Broad-band optical response at ambient pressure.}
In order to further understand the mechanism of the pressure-induced color change, we examine the optical response of \LHN\ in a broad frequency range. Figure~2a displays $R(\omega)$ in the frequency range of 100--50\,000~\icm\ at ambient pressure and temperature. The low-frequency $R(\omega)$ attains a high value and exhibits Hagen-Rubens [$(1-R) \propto \sqrt{\omega}$] behavior; a sharp plasma edge occurs at $\sim$12\,000~\icm. All these features demonstrate the metallic nature of \LHN. Two humps are observed in $R(\omega)$ at about 24\,000 and 40\,000~\icm, respectively, which are associated with inter-band electronic transitions. The Kramers-Kronig analysis of the measured $R(\omega)$ yields the real part of the optical conductivity $\sigma_{1}(\omega)$ which is directly related to the joint density of states and reflects the electronic band structure of the material~\cite{Basov2005RMP,Yang2020PRL,Jiang2020PRB,Hao2021PRB}. Figure~2b depicts $\sigma_{1}(\omega)$ of \LHN\ (cyan open circles) at ambient pressure. The low-frequency $\sigma_{1}(\omega)$ is dominated by a conspicuous Drude response, in accord with the metallic behavior of this material~\cite{Dasenbrock-Gammon2023Nature,Shan2023CPL,Ming2023arXiv,Xing2023arXiv,Cai2023arXiv}. Above 15\,000~\icm, $\sigma_{1}(\omega)$ is characterized by two hump-like features peaking at about 24\,000 (3~eV) and 40\,000~\icm\ (5~eV), respectively.

We fit the measured $\sigma_{1}(\omega)$ of \LHN\ to the Drude-Lorentz model,
%
%
\begin{equation}
\sigma_{1}(\omega) = \frac{2\pi}{Z_{0}} \left[\frac{\omega^{2}_{p,D}}{\tau(\omega^{2}+\tau^{-2})}
   + \sum_{i} \frac{\gamma_{i} \omega^{2} \omega_{p,i}^{2}}{(\omega_{0,i}^{2} - \omega^{2})^{2} + \gamma_{i}^{2} \omega^{2}}\right],
\label{DLModel}
\end{equation}
where $Z_{0} \simeq 377$~$\Omega$ is the vacuum impedance. The first term refers to a Drude profile which describes the optical response of free carriers or intra-band transitions and is characterized by a plasma frequency $\omega_{p,D}$ and a quasiparticle scattering rate $1/\tau$. The square of plasma frequency corresponds to the weight (integral) of the Drude profile: $\omega_{p,D}^{2} = Z_{0}ne^{2}/2\pi m^{\ast} = \frac{Z_{0}}{\pi^2}\int^{\infty}_{0}\sigma_{1}^{\text{Drude}}(\omega)d\omega$, where $n$ and $m^{\ast}$ are the carrier concentration and effective mass, respectively; $\omega^{2}_{p,D}$ also corresponds to the strength of intra-band transitions. The second term represents a sum of Lorentzian oscillators that are used to model inter-band transitions. In the Lorentz term, $\omega_{0,i}$, $\gamma_{i}$, and $\omega_{p,i}$ are the resonance frequency (position), damping (line width), and plasma frequency (strength) of the $i$th excitation.

As shown in Fig.~2b, the measured $\sigma_{1}(\omega)$ of \LHN\ can be described reasonably well by three components: a single Drude component (red hatched area) plus two Lorentz components L1 (blue hatched area) and L2 (green hatched area). The origin of these components can be elucidated through a comparison between our optical results and the calculated electronic band structure~\cite{Xie2023CPL}. Figure~4e displays the band structure of LuH$_{2}$ taken from Ref.~\cite{Xie2023CPL}. The blue and red lines denote the electronic bands at 0 and 10~GPa, respectively. Since the Drude component describes the optical response of free carriers or intra-band transitions, we attribute it to intra-band transitions within the bands crossing \EF\ near the $X$, $K$ and $U$ points in the Brillouin zone. The plasma frequency of the Drude component $\omega_{p,D}$ = 30950~\icm; assuming the effective mass $m^{\ast} = m_{e}$, this value corresponds to a carrier density of 1.06$\times$10$^{22}$~cm$^{-3}$, on the order of the carrier density for good metals. According to the positions of L1 (3~eV) and L2 (5~eV), it is reasonable to associate L1 with the inter-band transitions at the $\Gamma$ and $W$ points as indicated by the blue arrows in Fig.~4e, and ascribe L2 to the inter-band transitions at the $L$ point (green arrow in Fig.~4e). The same components also produce a reasonably good fit to the measured $R(\omega)$ of \LHN\ as shown by the dashed line in Fig.~2a. The plasma edge in $R(\omega)$ stems from free carriers (Drude response)~\cite{Dressel2002,Tanner2019}, and its position is given by $\omega_{p,D}/\sqrt{\epsilon_{c}}$, where $\epsilon_{c}$ is a constant representing the contribution of tightly-bound core electrons to the dielectric function~\cite{Tanner2019}.

\subsubsection*{Evolution of the optical response with pressure.}
Having understood the optical response of \LHN\ at ambient pressure, we now proceed to the analysis of the pressure dependence of $R(\omega)$. Based on the Drude-Lorentz fitting results at ambient pressure, we try to fit $R(\omega)$ at different pressures. Interestingly, we notice that only increasing the plasma frequency of the Drude component $\omega_{p,D}$ does not result in a satisfactory fit to the measured $R(\omega)$ under pressure. As shown by the blue dashed curves in Fig.~3a which are obtained by only increasing $\omega_{p,D}$, while the blue shift of the plasma edge in $R(\omega)$ can be qualitatively reproduced, the suppression of the high-frequency $R(\omega)$ is not captured by the fit. In order to describe the decrease of the high-frequency $R(\omega)$, the strength of L1 $\omega_{p,L1}$ must be reduced, suggesting that inter-band transitions are also involved in the pressure-induced color change in \LHN. If we allow both the Drude and L1 components to vary, a satisfactory fit is achieved for all pressures, as shown by the green dashed lines in Fig.~3b which reproduce not only the blue shift of the plasma edge in $R(\omega)$ but also the decrease of the high-frequency $R(\omega)$. The color of the sample can be reconstructed from the fitting $R(\omega)$ spectra at different pressures (see Supplementary Information). The sample color at different pressures in Fig.~3c and Fig.~3d is obtained from the fitting $R(\omega)$ curves (dashed lines) in Fig.~3a and Fig.~3b, respectively. It is obvious that the pressure dependence of the sample color in Fig.~3d, which is derived from the fitting $R(\omega)$ curves in Fig.~3b, agrees better with the experimental results in Fig.~1a, confirming that both the intra- and inter-band electronic transitions play an important role in the pressure-induced color change in \LHN.

The Drude-Lorentz fit of $R(\omega)$ at different pressures returns the evolution of the parameters with pressure (Fig.~4a--d). Figure~4a demonstrates an increase of the Drude weight $\omega_{p,D}^{2}$ with increasing pressure, pointing to an enhancement of intra-band transitions or an increase of free carrier concentration. The weight of L1 $\omega_{p,L1}^{2}$ decreases as the pressure rises (Fig.~4b), indicating that inter-band electronic transitions are suppressed by pressure. Figure~4c shows that the total weight of the Drude and L1 components $\omega_{p,D}^{2}+\omega_{p,L1}^{2}$ does not vary with pressure, suggesting a spectral weight transfer from L1 to the Drude component. Such a spectral weight transfer further implies that the application of pressure transforms some localized carriers into free carriers (Drude) in \LHN. In addition, as shown in Fig.~4d, the resonance frequency $\omega_{0,L1}$ of L1 increases with increasing pressure, implying that the interband transitions represented by L1 shift to higher energy.

All the above observations can be interpreted by the evolution of the band structure with pressure from first-principles calculations~\cite{Xie2023CPL}. As shown in Fig.~4e, at 0~GPa (blue lines), the electronic band structure of LuH$_{2}$ is characterized by (i) several bands crossing \EF\ near the $X$, $K$, and $U$ points, (ii) a hole-like valence band with the band top below but very close to \EF\ at the $\Gamma$ point, and (iii) an electron-like conduction band lying at about 2.5~eV above \EF\ at the $\Gamma$ point. Upon applying pressure, for example $p$ = 10~GPa (red lines), both the hole-like valence band and the electron-like conduction band at the $\Gamma$ point shift upwards, and the hole-like valence band crosses \EF. Since these two bands and their pressure dependence are essential for understanding the optical response of \LHN\ under pressure, we plot a schematic band structure of LuH$_{2}$ near the $\Gamma$ point in Fig.~4f based on the calculated band structure from Ref.~\cite{Xie2023CPL}. At ambient pressure, as shown by the blue curves, no band crosses \EF. Because the valence band is fully occupied and the conduction band is completely empty, no intra-band transitions are allowed near the $\Gamma$ point. Therefore, only inter-band transitions, as indicated by the blue solid arrow in Fig.~4f, can occur near the $\Gamma$ point. These inter-band transitions contribute partially to the weight of the L1 component. At high pressure, as illustrated by the red curves in Fig.~4f, both the valence and conduction bands shift upwards, and the top of the valence band is now above \EF. The presence of unoccupied states near the top of the valence band allows intra-band transitions to happen between the occupied (below \EF) and unoccupied (above \EF) states in the valence band. The occurrence of intra-band transitions or a Fermi surface (free carriers) near the $\Gamma$ point naturally explains the increase of $\omega_{p,D}^{2}$ with increasing pressure (Fig.~4a). Due to the depopulation of the valence band top, the inter-band transitions near the $\Gamma$ point, as denoted by the red dashed arrow, are now forbidden. This well explains the decrease of $\omega_{p,L1}^{2}$ with increasing pressure (Fig.~4b). Moreover, since the top of the valence band is depopulated at high pressure, the allowed inter-band transitions move away from the $\Gamma$ point, as indicated by the red solid arrow in Fig.~4f. Note that while the energy of the inter-band transitions at ambient pressure (blue solid arrow) equals to the band gap, the energy of the allowed inter-band transitions at high pressure (red solid arrow) becomes larger than the band gap as they move away from the $\Gamma$ point, providing a reasonable interpretation for the increase of $\omega_{0,L1}$ with increasing pressure (Fig.~4d).

%
%
\section*{Discussion}
Based on the above results, the pressure-induced color change in \LHN\ is readily comprehensible. We propose that the color tunability through pressure in \LHN\ relies on the coexistence of several deep bands crossing \EF\ near the $X$, $K$, and $U$ points, and a shallow valence band near the $\Gamma$ point whose band top is in the proximity of \EF. While the deep bands contribute appropriate carrier density which produces a sharp plasma edge in $R(\omega)$ right in the visible range, the shallow band enables the tunability of the optical response, because intra- and inter-band electronic transitions involving shallow bands are very sensitive to the band shift induced by external perturbations such as pressure. The application of pressure shifts the shallow valence band at the $\Gamma$ point from completely below \EF\ to partially above \EF, which results in an increase of intra-band transitions alongside a decrease of inter-band transitions near the $\Gamma$ point. Such a transformation between intra- and inter-band transitions increases the Drude weight $\omega_{p,D}^{2}$ but reduces the weight of L1 $\omega_{p,L1}^{2}$. The increase of $\omega_{p,D}^{2}$ gives rise to the blue shift of the plasma edge in $R(\omega)$, leading to the increase of $R_{\text{red}}$; The reduction of $\omega_{p,L1}^{2}$ causes the suppression of the high-frequency $R(\omega)$, resulting in the decrease of $R_{\text{blue}}$. The combination of the increase in $R_{\text{red}}$ and the decrease in $R_{\text{blue}}$ is responsible for the pressure-induced color change of \LHN.

Previous studies have reported different thresholds for the pressure-induced color change in different samples~\cite{Dasenbrock-Gammon2023Nature,Zhang2023arXiv,Shan2023CPL,Xing2023arXiv,Zhao2023arXiv}, which have been attributed to the difference in the content of N or H; recent theoretical calculations~\cite{Tao2023arXiv,Kim2023arXiv} have also demonstrated that the concentration of H vacancies has a strong influence on the sequence and pressure threshold of the color change in the LuH$_{2}$ and Lu-H-N systems. In this context, it is constructive to discuss the role of H in the pressure-induced color change. Hydrogenation may result in a compound with a large bulk modulus $K_{0}$, in which pressure is expected to create a large volume compression and a marked band shift, leading to a pressure-induced color change. However, $K_{0}$ for the Lu-H-N compound is about 88.6~GPa~\cite{Dasenbrock-Gammon2023Nature} which is even smaller than the value for gold (172.5~GPa)~\cite{Dewaele2004PRB}. Note that gold has a sharp plasma edge in the visible range (2.23~eV, $\lambda$ = 556~nm, yellow light)~\cite{Cooper1965PR}, but does not exhibit a noticeable color change under pressure. These facts indicate that a large $K_{0}$ and a plasma edge in the visible range are not sufficient to generate color tunability; a shallow band is indispensable for the pressure-induced color change. In this framework, a difference in the content of H or N in the Lu-H-N compound is expected to cause a shift of \EF. Such an \EF\ shift may affect not only the position of the plasma edge in $R(\omega)$, but also the energy of the shallow band with respected to \EF, thus resulting in different pressure thresholds for the color change in different samples.

To summarize, we investigated the optical response of LuH$_{2 \pm x}$N$_{y}$ in a broad frequency range at ambient pressure and the evolution of $R(\omega)$ with pressure in the visible spectral range. The broad-band optical spectra at ambient pressure reveal a Drude component associated with intra-band electronic transitions plus two Lorentz components (L1 and L2) arising from inter-band electronic transitions. As the pressure rises, the spectral weight is transferred from L1 to the Drude component, resulting in a blue shift of the plasma edge in $R(\omega)$ and a reduction of the high-frequency $R(\omega)$. These observations suggest that the transformation between intra- and inter-band electronic transitions may be responsible for the pressure-induced color change in LuH$_{2 \pm x}$N$_{y}$. Our results shed new light on the mechanism of the pressure-induced color change in the Lu-H-N system.

\section*{Methods}

\footnotesize{
\subsubsection*{Sample synthesis}
The detailed synthesis procedures of \LHN\ crystal can be found in Refs.~\cite{Ming2023arXiv,Zhang2023arXiv}. We synthesized the LuH$_{2\pm x}$N$_{y}$ samples under 300$^{\circ}$C and 2~GPa for 10~h using piston-cylinder type high pressure apparatus (LP 1000-540/50, Max Voggenreiter). The initial molecular ratio of Lu and N was 3:1. A mixture of NH$_{4}$Cl (Alfa Aesar 99.99\%) and CaH$_{2}$ (Alfa Aesar 98\%) in a molar ratio of 2:8 was used as the source of H$_{2}$ and NH$_{3}$.

\subsubsection*{Optical measurements}
In order to perform optical measurements, we polished the \LHN\ crystal with diamond lapping films to obtain a mirror-like surface. The near-normal-incidence reflectivity $R(\omega)$ at ambient pressure was measured in the frequency range of 100--50\,000~\icm\ using a Bruker Vertex 80v Fourier transform infrared spectroscopy (FTIR). An \emph{in situ} gold/silver evaporation technique~\cite{Homes1993} was adopted. $R(\omega)$ at different pressures from 0.8 to 33.2~GPa was measured in the visible range (13\,000--22\,000~\icm) using an Andor grating spectrometer (SR550i) in conjunction with a Nikon inverted microscope (Elipse Ti-U). Incident light was focus on the sample by a 50X objective, and the spot size on the sample can be controlled by an aperture in the microscope, which was adjusted smaller than the sample size. A symmetric diamond anvil cell (DAC) was used to generate the pressure on our sample. The pressure was calibrated using the ruby fluorescence method.

\subsubsection*{Kramers-Kronig analysis}
The real part of the optical conductivity $\sigma_{1}(\omega)$ was determined via a Kramers-Kronig analysis of the measured $R(\omega)$ for \LHN~\cite{Dressel2002,Tanner2019}. Below the lowest measured frequency (100~\icm), a Hagen-Rubens ($R = 1 - A\sqrt{\omega}$) form was used for the low-frequency extrapolation. Above the highest measured frequency (50\,000~\icm), we assumed a constant reflectivity up to 12.5~eV, followed by a free-electron ($\omega^{-4}$) response.

\subsection*{Data availability}
All data that support the findings of this study are available from the corresponding authors upon request.
}




{\footnotesize
\subsubsection*{Acknowledgements}
We thank Bing Xu, Huan Yang, and Run Yang for helpful discussions. We gratefully acknowledge financial support from the National Key R\&D Program of China (Grants No. 2022YFA1403201, and No. 2022YFA1404700), the National Natural Science Foundation of China (Grants No. 12174180, No. 12074085, and No. 12061131001), the Fundamental Research Funds for the Central Universities (Grant No. 020414380095), Jiangsu shuangchuang program.

\subsubsection*{Author contributions}
Z.L., Y.Z. and S.H. performed the ambient- and high-pressure optical measurements with the assistance of C.P., Q.L., X.Z., Y.D. and H.Y.; X.M. and X.Z. synthesized the crystals; Z.L., X.Z., Y.D., H.Y. and H.H.W. analyzed the data; Z.L., Y.D. and H.H.W. wrote the manuscript; all authors made comments on the manuscript; H.H.W. supervised the project.

\subsubsection*{Additional information}

\begin{trivlist}
\item[]\textbf{Supplementary information} accompanies this paper at ...

\item[]\textbf{Competing financial interests:} The authors declare no competing financial interests.

\item[]\textbf{Reprints and permission} information is available online.
\end{trivlist}
}


%
%
\clearpage
\centerline{\includegraphics[width=0.7\columnwidth]{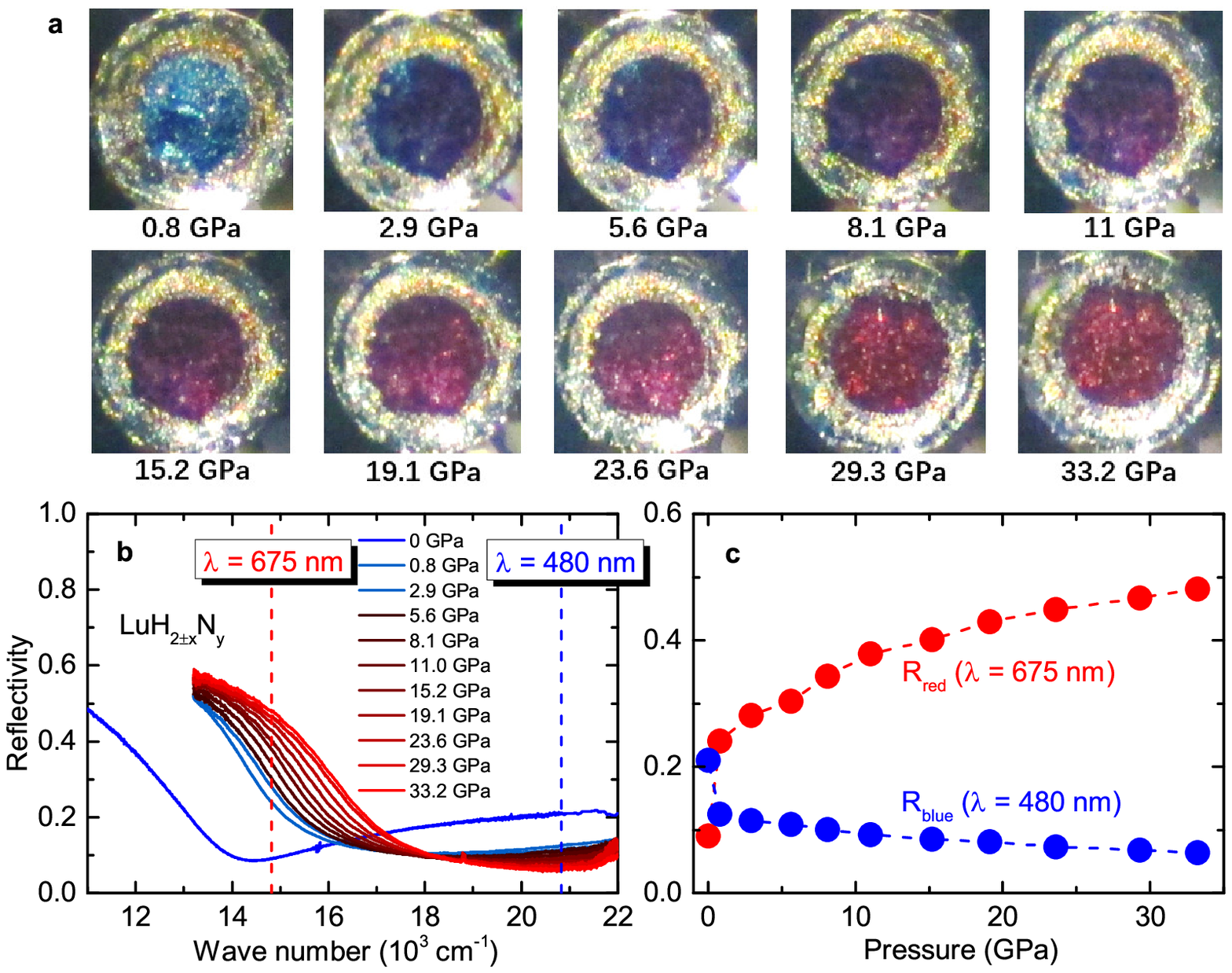}}%
\vspace*{0.8cm}
\noindent{\textsf{\textbf{Figure 1 $|$ Pressure-induced color change in LuH$_{2\pm x}$N$_{y}$.} \textbf{a} The optical microscope images of LuH$_{2\pm x}$N$_{y}$ at different pressures, manifesting a color change from blue to red. \textbf{b} The reflectivity $R(\omega)$ of LuH$_{2\pm x}$N$_{y}$ in the visible range at different pressures. \textbf{c} The evolution of $R(\omega)$ with pressure for red light $R_{\text{red}}$ (red solid circles) and that for blue light $R_{\text{blue}}$ (blue solid circles).}}
%
%
%
\clearpage
\centerline{\includegraphics[width=0.8\columnwidth]{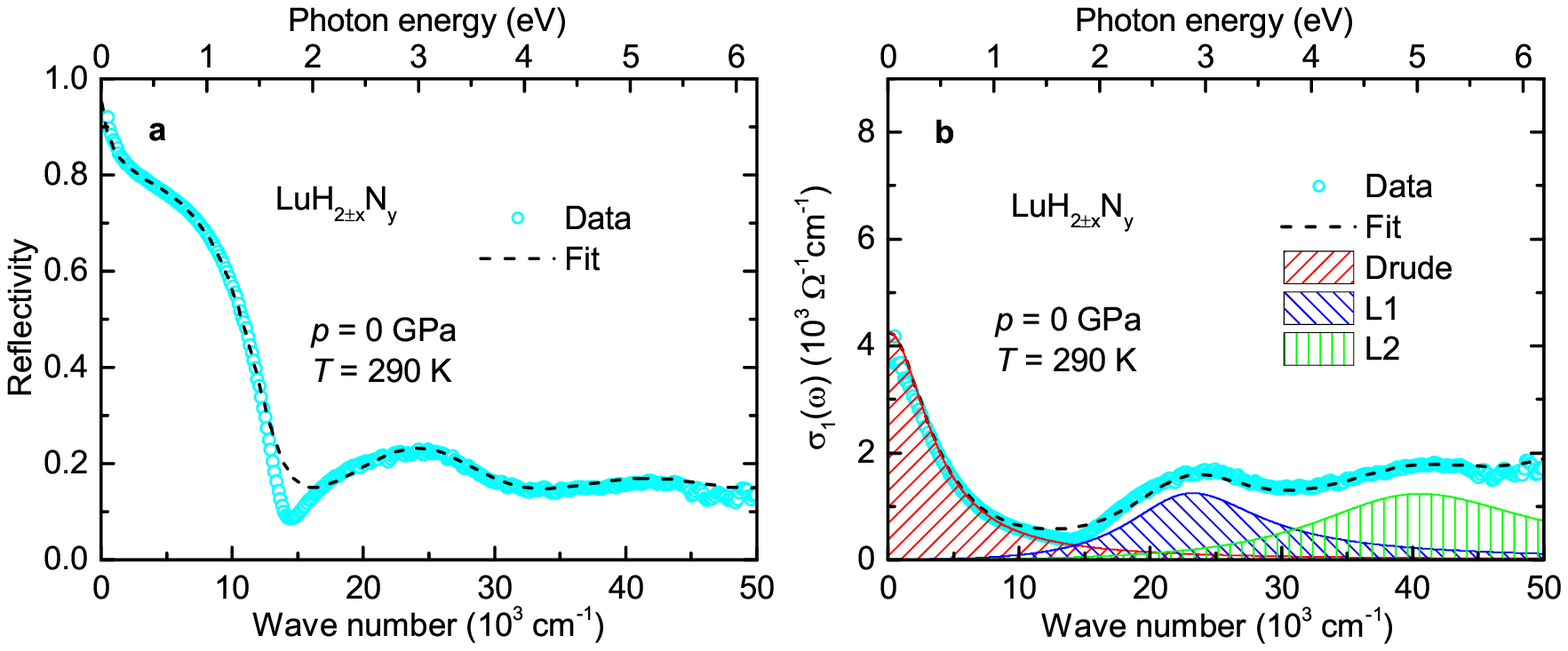}}%
\vspace*{0.8cm}
\noindent{\textsf{\textbf{Figure 2 $|$ Reflectivity and optical conductivity of LuH$_{2\pm x}$N$_{y}$.} \textbf{a} The reflectivity $R(\omega)$ of LuH$_{2\pm x}$N$_{y}$ in the frequency range 100--50\,000~\icm\ at ambient pressure. The dashed line represents the fitting result. \textbf{b} The optical conductivity $\sigma_{1}(\omega)$ of LuH$_{2\pm x}$N$_{y}$ measured at ambient pressure. The dashed line through the data is the Drude-Lorentz fit, which is decomposed into a Drude component (red hatched area) plus two Lorentz components L1 (blue hatched area) and L2 (green hatched area).}}

%
%
\clearpage
\centerline{\includegraphics[width=\columnwidth]{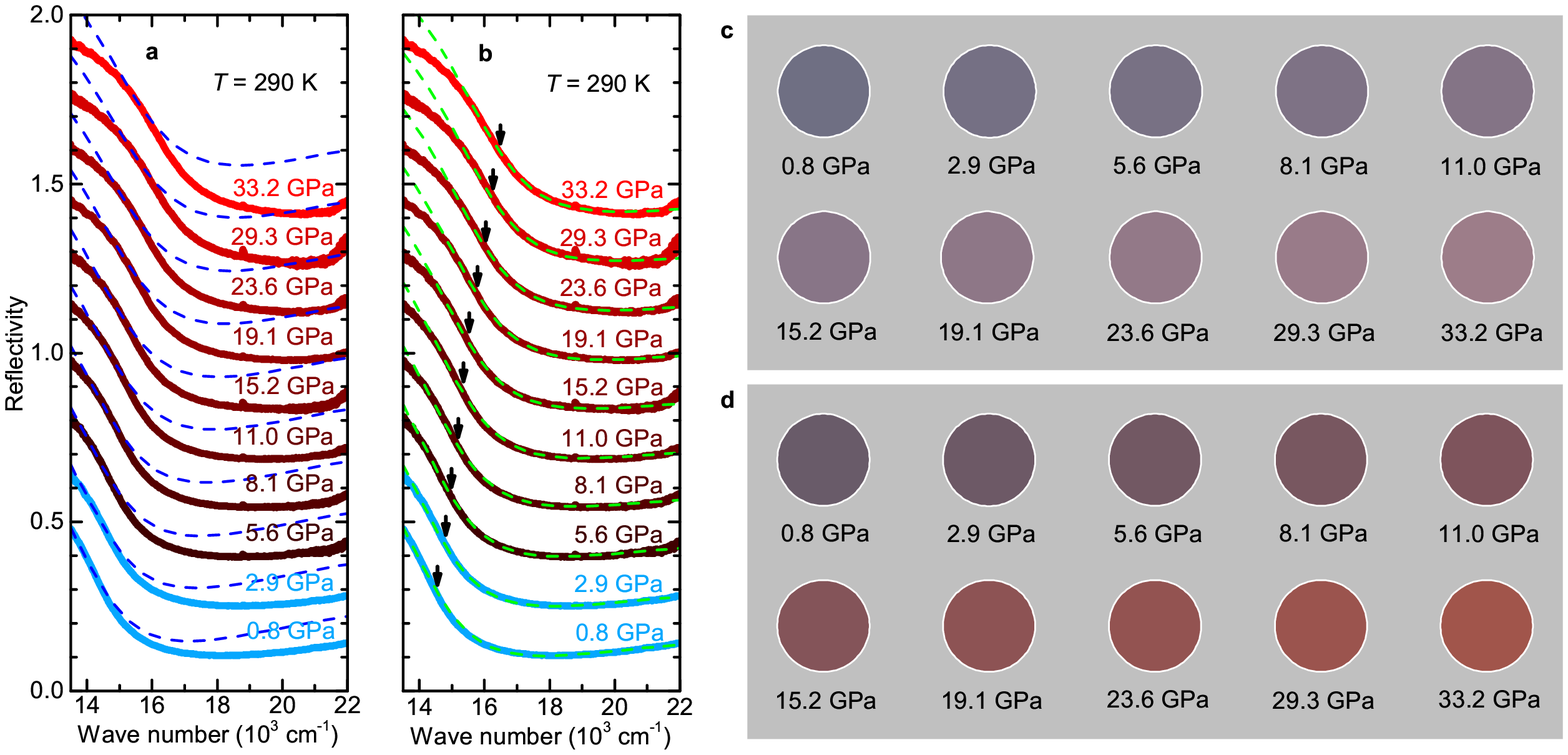}}%
\vspace*{0.8cm}
\noindent{\textsf{\textbf{Figure 3 $|$ Modeling the pressure dependence of $R(\omega)$ and the color change.} \textbf{a}, \textbf{b} The thick solid lines are $R(\omega)$ of \LHN\ in the visible range at different pressures. The blue dashed lines are the Drude-Lorentz fit obtained by only increasing the Drude plasma frequency. The green dashed lines represent the fit obtained by allowing both the Drude plasma frequency and L1 to vary. The black arrows indicate the position of the plasma edge in $R(\omega)$ which shifts to higher frequencies as the pressure increases. \textbf{c}, \textbf{d} The color of the sample at different pressures calculated from the fitting $R(\omega)$ in \textbf{a} and \textbf{b}, respectively.}}

%
%
\clearpage
\centerline{\includegraphics[width=0.9\columnwidth]{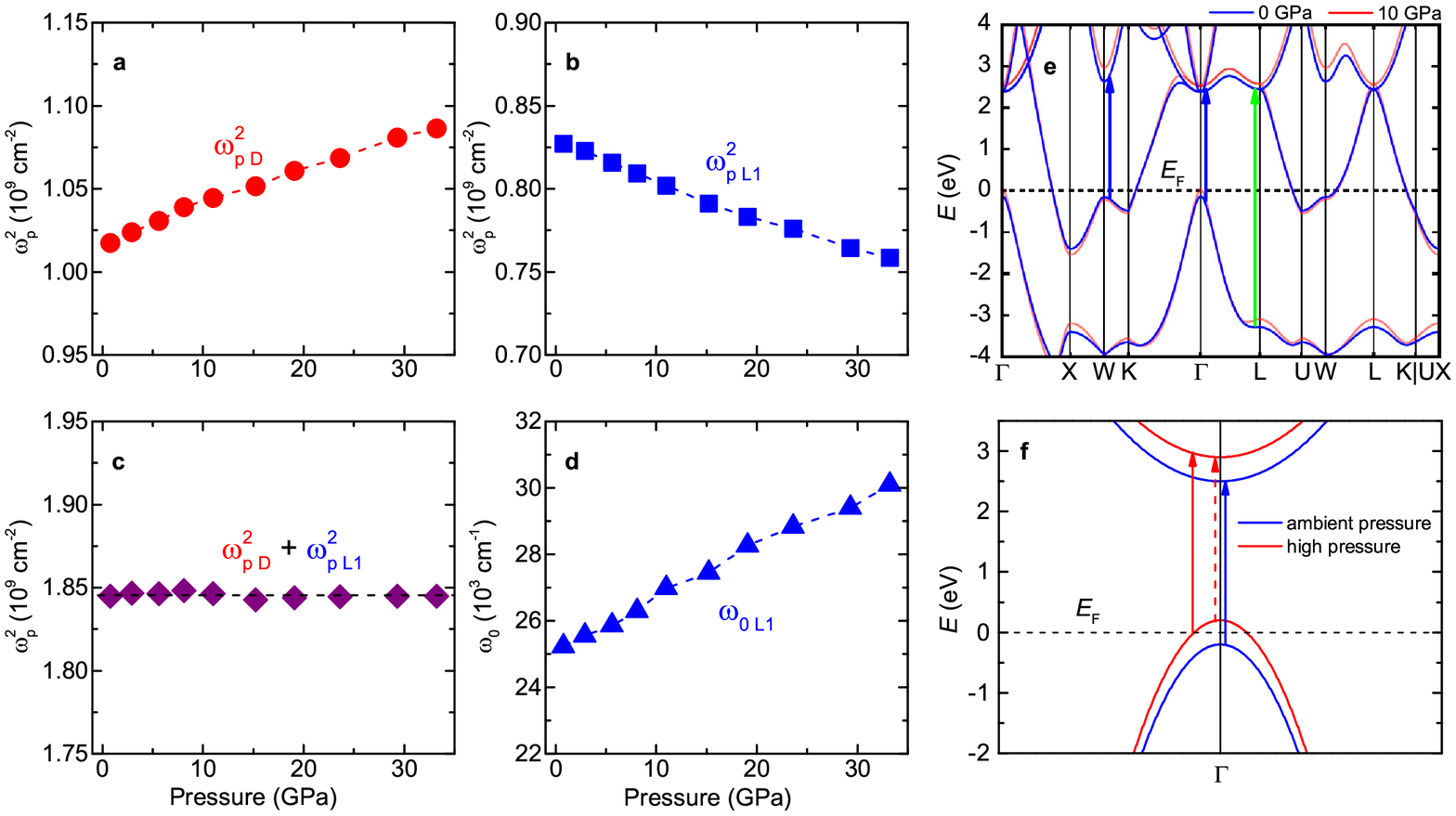}}%
\vspace*{0.8cm}
\noindent{\textsf{\textbf{Figure 4 $|$ Pressure dependence of fitting parameters and schematic band structure.} The pressure dependence of the Drude weight $\omega_{p,D}^{2}$ (\textbf{a}), the weight of L1 $\omega_{p,L1}^{2}$ (\textbf{b}), the sum of $\omega_{p,D}^{2}$ and $\omega_{p,L1}^{2}$ (\textbf{c}), and the resonance frequency of L1 $\omega_{0,L1}$ (\textbf{d}). \textbf{e} The calculated electronic band structure of LuH$_{2}$ taken from Ref.~\cite{Xie2023CPL}. The blue and red lines denote the electronic bands at 0 and 10~GPa, respectively. The arrows denote inter-band electronic transitions. \textbf{f} The schematic electronic structure near the $\Gamma$ point for LuH$_{2}$ at ambient (blue curves) and high (red curves) pressure. The solid arrows denote inter-band electronic transitions; the dashed arrow indicates inter-band transitions that can not happen.}}


\end{document}